\documentclass[%
 reprint,
 amsmath,amssymb,
 aps,
]{revtex4-2}

\usepackage{graphicx}
\usepackage{dcolumn}
\usepackage{bm}
\usepackage{amsmath}
\usepackage{lipsum}

\begin{document}

\preprint{APS/123-QED}

\title{Extending depolarized DLS measurements to turbid samples}

\author{Antara Pal}
 \affiliation{Division of Physical Chemistry, Department of Chemistry, Lund University, Lund, Sweden.} 
\author{Peter Holmqvist}%

\affiliation{Division of Physical Chemistry, Department of Chemistry, Lund University, Lund, Sweden.}



\author{Andrea Vaccaro }
\affiliation{LS Instruments, Fribourg, Switzerland.}
\author{Peter Schurtenberger}
 \email{peter.schurtenberger@fkem1.lu.se}
\affiliation{Division of Physical Chemistry, Department of Chemistry, Lund University, Lund, Sweden.}
\affiliation{Lund Institute of advanced Neutron and X-ray Science LINXS, Lund University, Lund, Sweden}


\date{\today}

\begin{abstract}
The application of dynamic light scattering to soft matter systems has strongly profited from advanced approaches such as the so-called modulated 3D cross correlation technique (mod3D-DLS) that suppress contributions from multiple scattering, and can therefore be used for the characterization of turbid samples. 
Here we now extend the possibilities of this technique to allow for depolarized light scattering (Mod3D-DDLS) and thus obtain information on both translational and rotational diffusion, which is important for the characterization of anisotropic particles. We describe the required optical design and test the performance of the approach for increasingly turbid samples using well defined anisotropic colloidal models systems. Our measurements demonstrate that 3D-DDLS experiments can be performed successfully for samples with a reduced transmission due to multiple scattering as low as 1\%. We compare the results from this approach with those obtained by standard DDLS experiments, and point out the importance of using an appropriate optical design when performing depolarized dynamic light scattering experiments with turbid systems.
\end{abstract}

\maketitle

\section*{Introduction}
Anisotropic building blocks with different shapes such as rods, cylinders and plates are frequently found in biological macro-molecules, micelles and other colloidal systems~\cite{dhont1996introduction, missel1983influence, neeson1983electric}. The shape of the building blocks affects not only the spatial arrangement for self-assembled structures but also the translational and rotational dynamics for these systems. For spherical colloids, their dynamic behavior is well understood, and a very good agreement has been established between theory, simulations, and experiments~\cite{banchio2018short, heinen2011pair, westermeier2012structure}. For anisotropic particles, well established theory and predictions can be found for dilute systems~\cite{ortega2003hydrodynamic, perrin1934brownian, perrin1936brownian, broersma1959diffusion}, which also have been successfully compared with experiments for both translational and rotational motions~\cite{gunther2011rotational, kleshchanok2012dynamics, martchenko2011hydrodynamic, nixon2019depolarized, rodriguez2007dynamic, feller2021translational} mainly using dynamic light scattering (DLS) and depolarized dynamic light scattering (DDLS)~\cite{berne2000dynamic}. DDLS is the most useful tool for characterizing nonspherical particles, and thus has been applied to optically anisotropic particles such as the tobacco mosaic virus~\cite{lehner2000determination}, gold nanorods~\cite{van2000colloidal}, spheres with internal optical anisotropy~\cite{koenderink2000rotational, geers2016new, balog2014dynamic, Balog2015}, carbon nanotubes~\cite{shetty2009multiangle}, claylike particles~\cite{kleshchanok2012dynamics, jabbari2004ageing} and cellulose nanocrystals~\cite{khouri2014determination}. However, all these measurements are limited to rather low concentrations, beyond which multiple scattering and or absorption by the samples play a crucial role to obtain reliable data.\par
 
Today there are several dynamic light scattering-based measurement techniques available to measure translational diffusion in turbid samples~\cite{medebach2007dynamic, overbeck1997probing, schatzel1991suppression, urban1998characterization, Block2010}. However, the rotational diffusion for opaque samples or samples suffering from multiple scattering is still somewhat inaccessible by the available measurement techniques. We have now implemented the depolarized measurement option on the modulated 3D cross correlation technique (mod3D) \cite{Block2010} in order to perform DDLS experiments with turbid samples to characterize the rotational motions of anisotropic colloids even in the multiple scattering regime. With this development one can now investigate the depolarized scattering on samples with a substantial amount of turbidity. This will not only greatly increase the concentration range at which these investigations can be performed, but also the different type of systems that can be measured, e.g. systems which are made up of inorganic materials.\par

In our case study we have investigated aqueous dispersions of colloidal hematite/silica core/shell ellipsoids with two different aspect rations at several different concentrations as a strongly scattering model system. We demonstrate the feasibility of 3D-DDLS measurements down to a transmission as small as 1\%. We highlight the importance of an approach capable of suppressing multiple scattering in DDLS by comparing our data with standard DDLS measurements and theoretical predictions. We also discuss the experimental set-up and the technical and experimental challenges that are posed by these measurements.\par

\section*{Experimental section}
\subsection*{Materials}
\subsection*{Synthesis and Transmission Electron Microscopy (TEM)}
Silica/hematite core/shell ellipsoidal particles of two different aspect ratios were synthesized in a two step process. First the hematite ellipsoids were synthesized in water following the approach described by Ocana et al.~\cite{ocana1999homogeneous}. They were then coated with silica layer(s) in ethanol using the method described by Graf et al.~\cite{graf2003general}. The aspect ratio of the particles was controlled by tuning the thickness of the silica shell. Particles were purified by repeated centrifugation/re-dispersion cycles in water and were kept in water as a stock dispersion.\par
\begin{figure}[hb]
	\centering
		\includegraphics[width=0.52\textwidth]{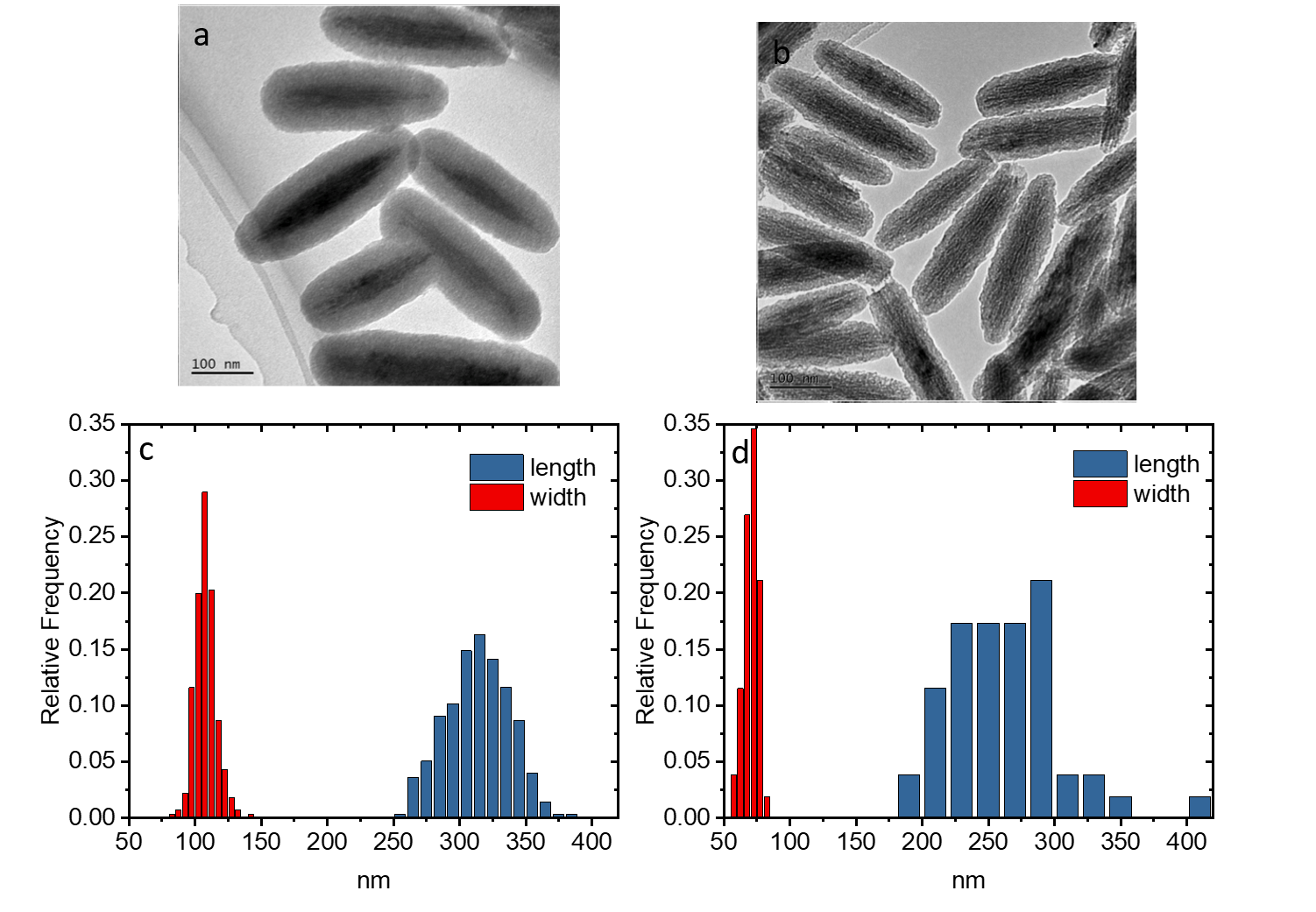}
	\caption{(a), (b) Representative TEM images for core-shell hematite-silica ellipsoidal particles of two different aspect ratios $\rho_1$ =2.9, and $\rho_2$=3.7, respectively. (c), (d) represent the statistical distribution of length scales as has been obtained from (a) and (b).}
	\label{fig:TEM}
\end{figure}
A transmission electron microscope (TEM) (JEOL 3000F microscope operating at 300 kV) was used to characterize both the size and shape of the particles. Particle size distributions were calculated by measuring at least 100 particles from TEM micrographs using the software ImageJ. For the batch of particles that we have named $\rho_1$, we found the long and short axes to be L$_1$ = 316 $\pm$ 26.3 nm and D$_1$ = 108 $\pm$ 7 nm respectively, leading to an aspect ratio of $\rho_1$ = 2.9, while for another batch, named $\rho_2$, L$_2$ = 266 $\pm$ 19 nm and D$_2$ = 72 $\pm$ 6 nm corresponding to $\rho_2$ = 3.7. Fig.~\ref{fig:TEM} (a) and (b) show representative TEM images for the ellipsoids and (c) and (d) shows the corresponding size distributions.
\subsection*{3D-DLS and 3D-DDLS}

\begin{figure}[h]
	\centering
		\includegraphics[width=0.48\textwidth]{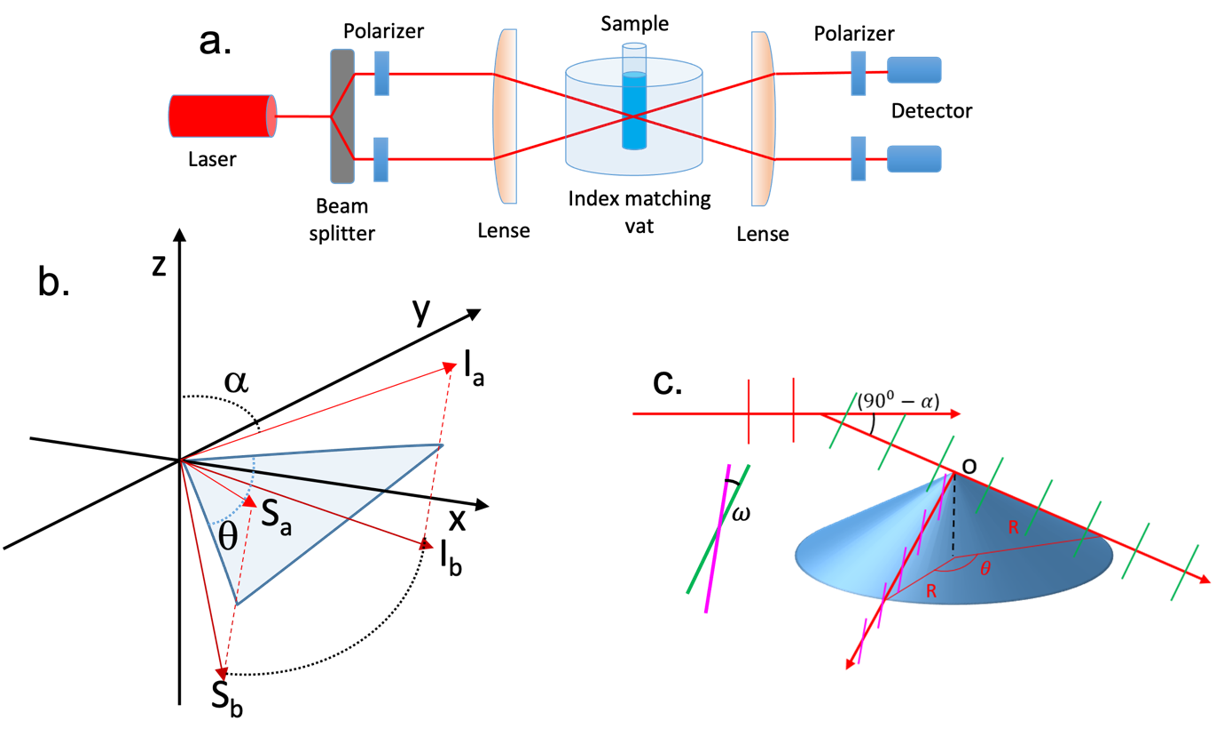}
		\includegraphics[width=0.40\textwidth]{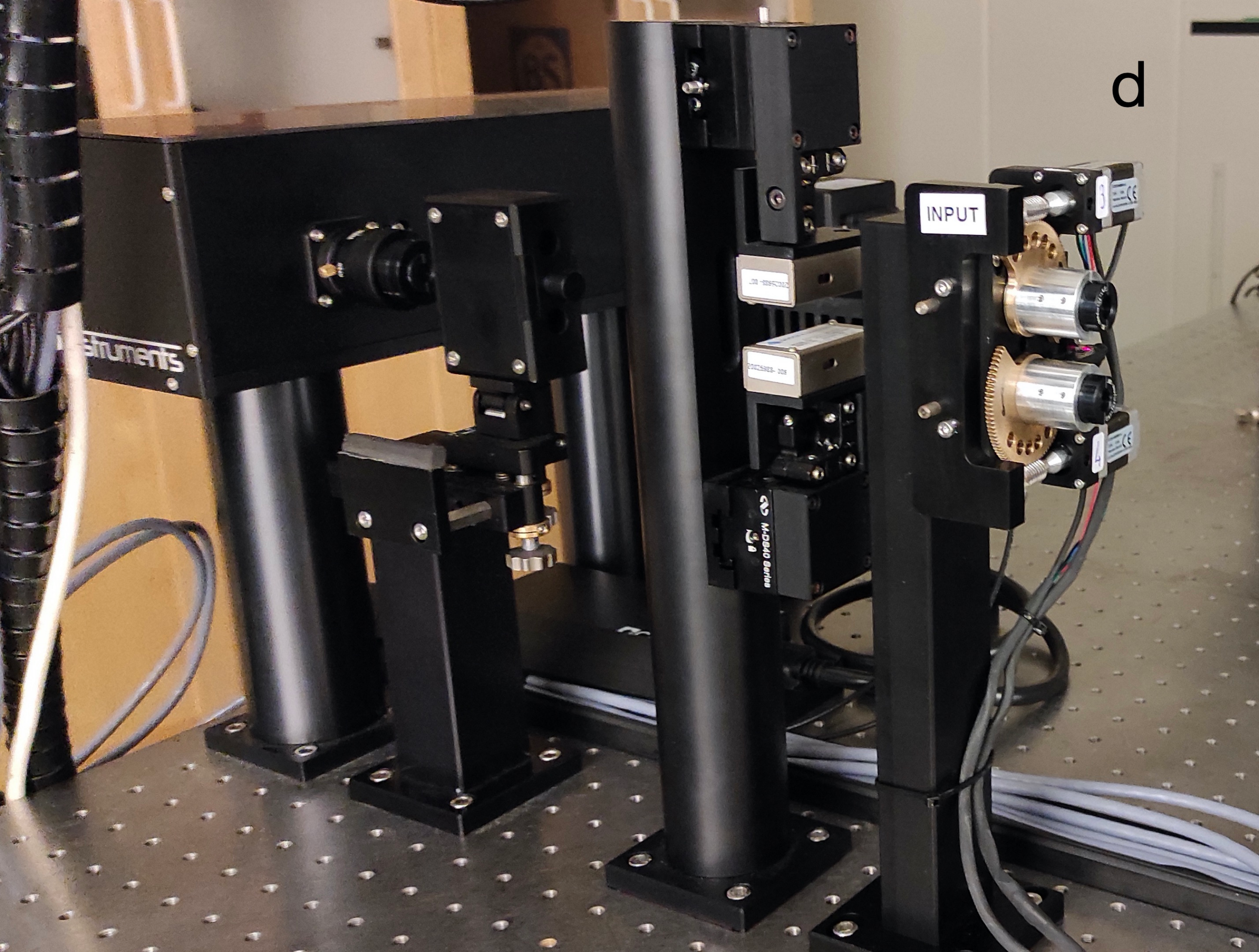}
	\caption{a) Schematic representation of the 3D-DLS set up with the addition of polarizers before and after the sample to enable 3D-DDLS measurements. b) The scattering geometry of the 3D set-up where I$_{a}$ and I$_{b}$ represent the vectors of the two incident beams and S$_{a}$ and S$_{b}$ scattering vectors at scattering angle $\theta$. The transparent triangle represents the mirroring plane between the two beams and is the plane at which a 2D measurement is performed. $\alpha$ is the angle between the normal of the mirroring plane and the incident beam and determine the angle of the scattering cone in 3D c) The scattering geometry of one of the beams in the scattering volume for the 3D-DLS set-up. $\alpha$ is the cone angle and $\omega$ is the turning of the polarization. d) Photograph of the actual motorized set-up at the beam splitter side of the 3DDLS set up.}
	\label{fig:schematic}
\end{figure}

The DLS measurements were performed on a 3D-DLS Spectrometer (LS Instruments, Switzerland) that implements the modulated 3D cross correlation technology \cite{Block2010}, equipped with 660 nm Cobolt laser with a maximum power of 100 mW. In the standard set up the measurements are performed in a VU polarization geometry where the single beam from the laser is vertically polarized. This means that the split beams are vertically (V) polarized for the incident beams in 3D geometry. There are no polarizers at the detector side giving an unpolarized (U) detector condition. In our modified set-up we have added two Glen-Thomson polarizers, with an extinction ratio larger than 10$^{5}$, and vertically(V) orientated them at the detector side, thus giving us a VV polarization geometry. For the 3D depolarized dynamic light scattering, 3D-DDLS, two additional identical Glen-Thomson polarizers, were mounted after the beam splitter in the 3D geometry (figure~\ref{fig:schematic}(a)). A VH geometry was then achieved by minimizing the detected scattering intensity of an isotropic sample containing spherical particles, prior to experiments at each angle. 5 mm cylindrical glass cells were used and placed in the temperature-controlled index matching bath containing decalin. The scattered light was detected within an angular range of 30 -- 135$^{\circ}$ by avalanche photodiodes and processed by an LS instrument correlator.
\subsection*{Polarization in 3D-DLS set-up}
The 3D cross-correlation technology uses two laser beams that cross each other within the sample as shown schematically in Fig.~\ref{fig:schematic}(a). This then allows for performing two independent DLS experiments in the same scattering volume using exactly the same scattering vector $\overrightarrow{q}$ both in magnitude and in direction (Fig.~\ref{fig:schematic}(b)). The intensities measured by the detectors of these two experiments are then cross-correlated, resulting in an effective suppression of all multiple scattering contributions. As a result, the measured cross-correlation function corresponds to the correct intermediate scattering function without multiple scattering contributions, while the multiply scattered light only results in a decrease in the intercept of the correlation function. In order to perform both standard 3D-DLS as well as 3D-DDLS experiments, polarizers need to be added accordingly as shown in Fig.~\ref{fig:schematic}(a).\par
However, due to the non-normal incidence of the incident and scattered beams, the combination of 3D cross correlation technology and DDLS requires a different instrument design. In order to perform accurate DDLS measurements the polarizers at the detectors have to be aligned such that they are perpendicular to the polarization of the incident beam. In standard 2D scattering geometry there is a single scattering plane defined by the incident beam and the scattered light, and the polarizer at the (single) detector needs to be turned by 90$^{\circ}$ away from the initial vertical polarization direction for all scattering angles in order to achieve the correct VH polarization geometry. In contrast, in 3D geometry, the incident beam points towards the sample at an angle (90$^{\circ}$ - $\alpha$), which makes the wave vector of the scattered light follow the surface of a cone (Fig.~\ref{fig:schematic}(c)) with an angle $\alpha$. In order to achieve a correct VH geometry, this then requires an adjustment of the direction of the polarizers in front of the detectors with a tilt angle that depends on the scattering angle $\theta$. This is illustrated in Fig.~\ref{fig:schematic}(c), where the direction of polarization for the scattering angle $\theta = 0^{\circ}$ is shown by green lines, whereas pink lines indicate the same for another scattering angle ($0^{\circ} < \theta < 180^{\circ}$). This change in the direction of polarization given by the tilt angle $\omega$ (as shown in Fig.~\ref{fig:schematic}(c) inset) can be derived analytically as a function of $\theta$. The corresponding expression for $\omega(\theta)$ is given by (detailed calculations are included in Appendix):
\begin{equation}
\omega = \cos^{-1}\left(\frac{\cos^{2}\alpha \;\cos\theta\;+\;\sin^{2}\alpha}{\sqrt{1-\sin^{2}\alpha\:\cos^{2}\alpha \:(1 - \cos\theta)^{2}}}\right)		
	\label{eq1}
\end{equation}
Here $\alpha$ is the angle of the scattering cone due to focusing lens (Fig.~\ref{fig:schematic}(a)). This renders DDLS experiments more demanding as the direction of the polarizers need to be adjusted at each scattering angle, and we have therefore designed a motorized set of holders that allow for high precision alignment of the polarizers according to the relation given by eqn. \ref{eq1} as illustrated in Fig.~\ref{fig:schematic}(d).

\begin{figure}[hbt]
	\centering
		\includegraphics[width=0.5\textwidth]{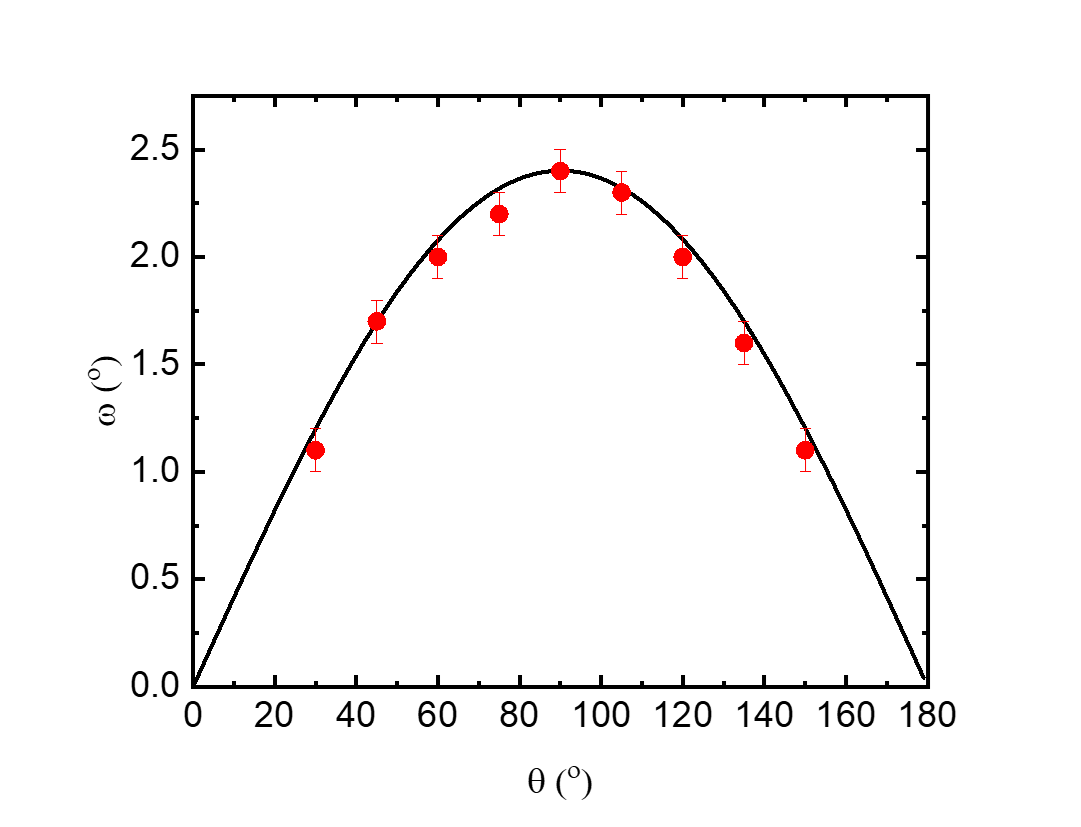}
	\caption{Shift of the polarizer, $\omega$, from classical cross polarization as a function of scattering angle, $\theta$.}
	\label{fig:polarization}
\end{figure}

\subsection*{Transmission measurements}
The transmission measurements were performed in 5 mm glass tubes using a 660 nm Cobolt laser of 100 mW. The transmission is defined as the ratio between the intensity measured after a sample with only solvent, $I_{0}$, and after a sample with particles, $I_{s}$, The transmission, T, is then defined as $T=I_{s}/I_{0}$.
\section*{Results}
To investigate the validity of eqn.~\ref{eq1}, an isotropic sample containing spherical particles was mounted and the incident beams were adjusted to be vertically polarized. The polarizers at the detectors were turned horizontally (90$^{\circ}$ in relation to the incident polarization) and then carefully readjusted from that position until the detected scattering intensity was minimized. In figure~\ref{fig:polarization} the experimentally observed tilt angle or rotation away from classical cross polarization, i.e. 90$^{\circ}$ degree turned in relation to the incident beam, is shown (red symbols) as a function of scattering angle $\theta$. The line in figure~\ref{fig:polarization} represents the theoretical calculation using eqn.~\ref{eq1} for an incident angle of the laser beam of $\alpha$ = 2.4$^{\circ}$. It can be noted that the polarization is not affected at 0$^{\circ}$ and 180$^{\circ}$ and the tilt angle $\omega$ is symmetric around 90$^{\circ}$ where it has its largest value of about 2.4$^{\circ}$. \par

This readjustment of the polarizer orientation has the same magnitude for the two different beams, but they should be turned in opposite directions: one turns clockwise while the other counterclockwise. 

\begin{figure}[h]
	\centering
		\includegraphics[width=0.48\textwidth]{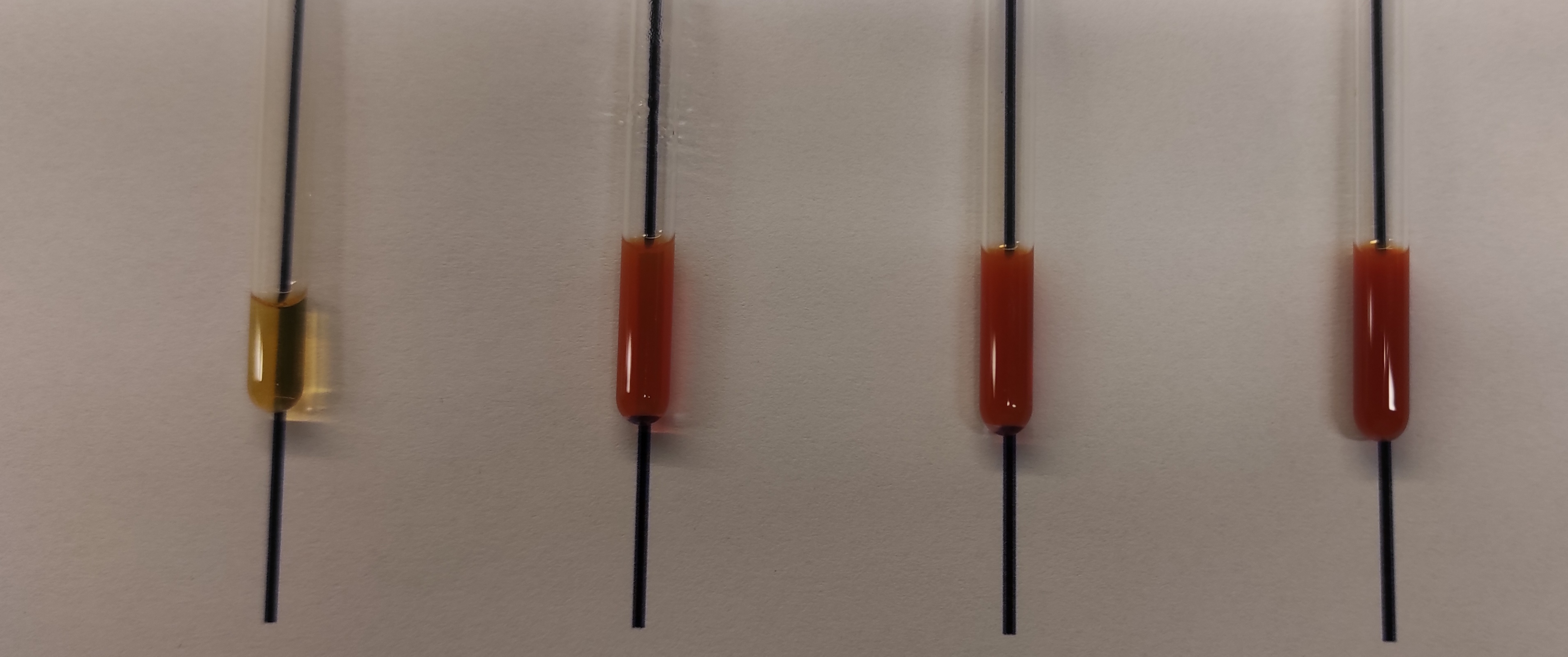}
	\caption{Photograph of aqueous suspensions of silica-coated hematite ellipsoids (axial ratio $\rho_2 = 3.7$) at 4 different concentrations: from left to right $c$ = 0.005, 0.05. 0.1, 0.2 wt$\%$.}
	\label{fig:samples}
\end{figure}

The performance of the 3D-DDLS implementation described in Fig.~\ref{fig:schematic} was investigated from a  series of test measurements with well-chosen model systems made by inorganic ellipsoidal colloids with very large scattering cross section that exhibit multiple scattering already at quite low concentrations, where interparticle interactions are still negligible. These particles are made up by a hematite core and an added silica shell which provides both improved colloidal stability and low polydispersity as well as allows for a convenient selection of the axial ratio. As described in the materials and methods section, two batches with different axial ratios were synthesized and characterized, and the details of their size distributions are given in Fig.~\ref{fig:TEM}.  A number of samples with different concentrations were then prepared, and 3D-DLS and 3D-DDLS measurements were performed to effectively suppress multiple scattering. Polarized ($g_{vv}(t)$) and depolarized ($g_{vh}(t))$ intensity crosscorrelation functions were obtained at different angles, and the corresponding translational and rotational diffusion coefficients were extracted from the angular dependence of $g_{vv}(t)$ and $g_{vh}(t)$, respectively.\par

Due to the high refractive index of the particles, the system gets turbid already at low colloidal concentrations. This is demonstrated in Fig.~\ref{fig:samples} with pictures from different samples in the 5 mm cylindrical glass cells at concentrations between 0.005 - 0.2 wt$\%$. The black line behind the cells is only visible for the lowest concentration, whereas the turbidity of the samples at larger concentrations makes them completely intransparent. From an optical inspection the particles were however well dispersed and did not sediment visibly during the measurements. All the samples were measured at scattering angles between 30$^{\circ}$ and 135$^{\circ}$ scattering angles in steps of 15$^{\circ}$. \par

\subsection*{3D-DLS measurements}

\begin{figure*}[hbt]
	\centering
		\includegraphics[width=0.8\textwidth]{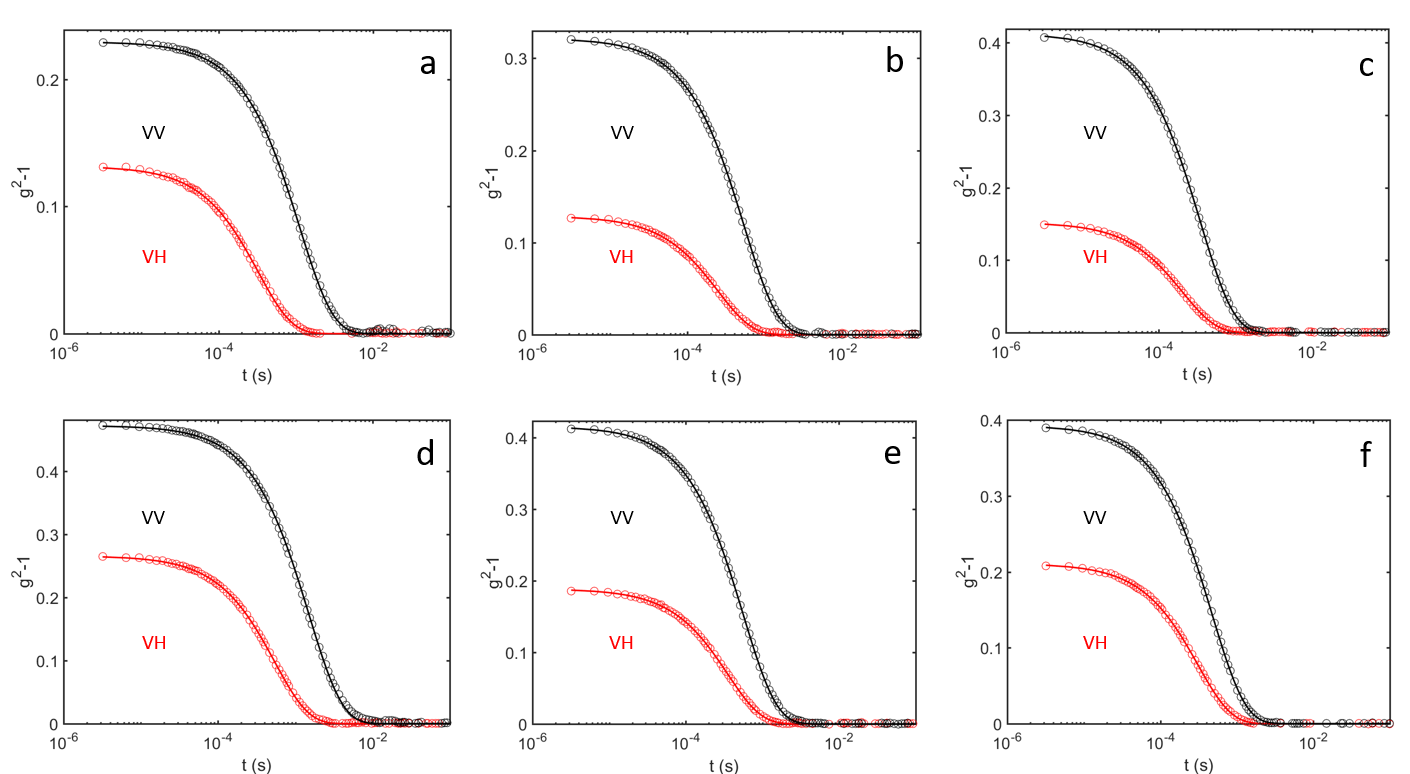}
	\caption{Normalized correlation functions, $g_{2}$ - 1, for VV (black symbols) and VH (red symbols) measurements for two different aspect ratios, $\rho_{2}$ = 3.7, concentration 0.05wt$\%$, transmission = 0.3540, (a-c) and $\rho_{1}$ = 2.9, concentration 0.05wt$\%$, transmission = 0.4145 (d-f). Three different scattering angles, $\theta$, are shown: 60$^{\circ}$ (a,d), 90$^{\circ}$ (b,e) and 120$^{\circ}$ (c,f).}
	\label{fig:correlation}
\end{figure*}

We first perform 3D-DLS measurements where the measured intensity cross-correlation function $g_{2}(q,t)$ corresponds to a VV geometry, i.e. we determine $g_{vv}(t)$. Measuring in this geometry at different angles $\theta$ would require in principle a change in the orientation of the polarizer in front of the detectors according to eqn.~\ref{eq1} and Fig.~\ref{fig:polarization}. While the polarizers can be adjusted automatically to the correct position according to eq. 1 when using motorized polarizers, this renders correct 3D-DLS more tedious if the polarizers need to be changed manually. However, performing these experiments with a constant orientation that corresponds to the one of the direction of polarization of the incident beam, i.e. a tilt angle $\omega = 0^{\circ}$, will have a very small effect on the detected scattered light. With a maximum change of the polarization of 2.4$^{\circ}$ the maximum loss of intensity if the polarizers are kept in the same orientation as the incident light is less than 0.1\%.  \par

Typical examples of the measured normalised intensity cross-correlation functions, $g_{2}(q,t) - 1$, in VV geometry are shown in Fig.~\ref{fig:correlation} for both aspect ratios $\rho_1$ and $\rho_2$ and different scattering angles. The correlation functions are well described by a single exponential:
\begin{equation}
	g_{2}(q,t) - 1 = f e^{-2\Gamma t}
	\label{eq2}
\end{equation}
where $f$ is the amplitude of the correlation function and $\Gamma$ is the relaxation rate of the density fluctuation (see black lines in Fig.~\ref{fig:correlation}). It is important to point out that in general $g_{vv}(t)$ is not characterized by a single exponential decay for optically anisotropic particles, since it contains also contributions from $g_{vh}(t)$ \cite{Degiorgio1994, balog2014dynamic}. For the present system the contribution from the depolarized scattering is more than 100 times smaller than the polarized part and cannot be seen in $g_{vv}(t)$. However, for particles with a much larger depolarization ratio such as gold nanoparticles the depolarized contribution can be seen also in the VV measurements, resulting in a double exponential decay of $g_{vv}(t)$~\cite{rodriguez2007dynamic, balog2014dynamic}.\par
\begin{figure}[hb]
	\centering
		\includegraphics[width=0.35\textwidth]{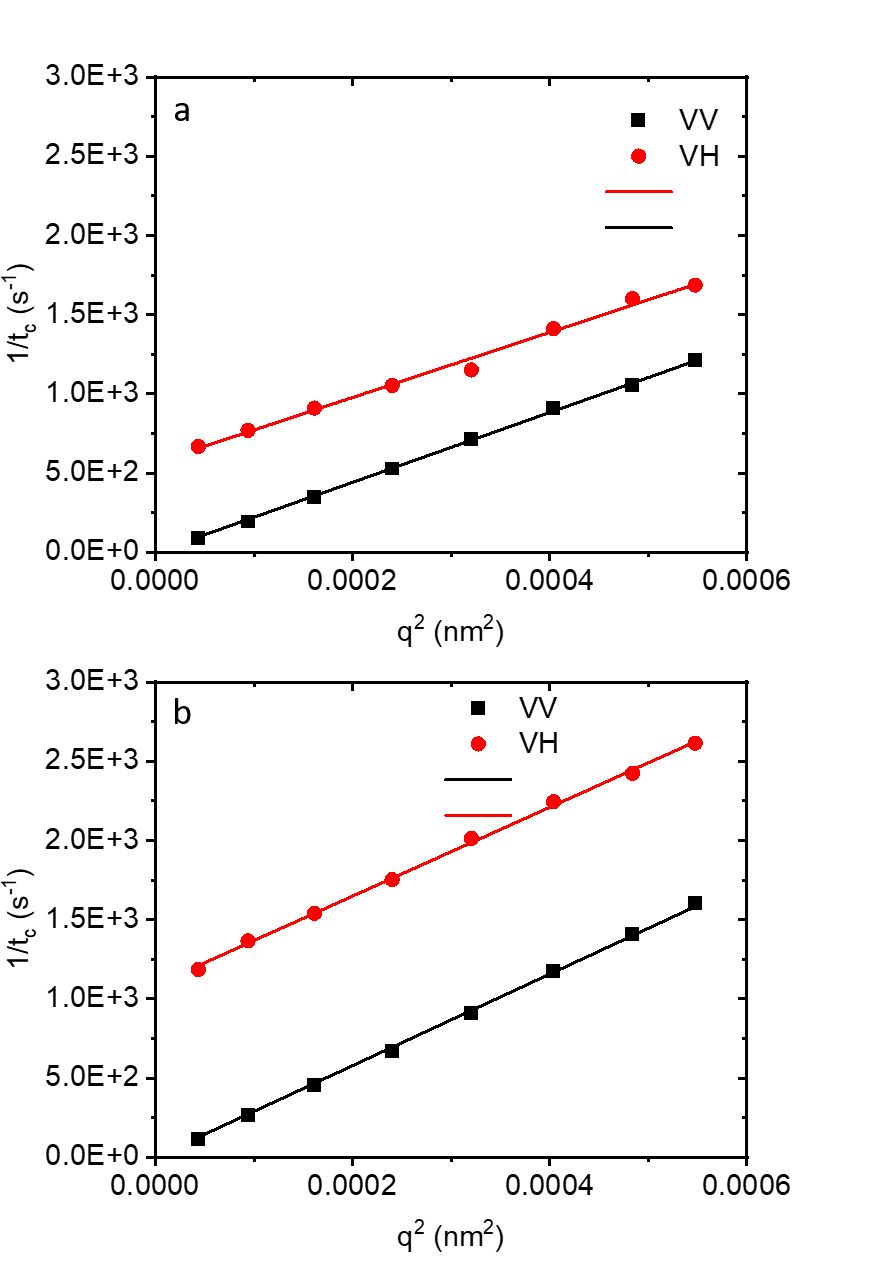}
	\caption{The extracted relaxation rate, $\Gamma$, as a function of q${^2}$ for VV (black) and VH (red) measurements for the two different aspect ratios, $\rho_{2}$ = 3.7 (a) and $\rho_{1}$ = 2.9 (b). Lines are fits to Eqn.~\ref{eq3} for VV (black) and Eqn.~\ref{eq4} for VH (red). Concentrations used were 0.05wt$\%$ for both samples.}
	\label{fig:Dt_Dr}
\end{figure}

Eqn.~\ref{eq2} was used to analyse the data obtained from the 3D-DLS measurements, and relaxation rates $\Gamma$ were extracted accordingly. The $q$-dependence of $\Gamma$ is given by
\begin{equation}
\Gamma = D_{T}\;q^{2}
	\label{eq3}
\end{equation}
Therefore the translational diffusion coefficient $D_{T}$ of the particles can be obtained from the angular dependence of $\Gamma$ for each sample, as it is directly related to the slope of the data plotted as $\Gamma$ vs q${^2}$. This is shown in Fig.~\ref{fig:Dt_Dr} for the two different aspect ratios at one concentration (0.05wt$\%$). \par 

\subsection*{3D-DDLS measurements}

In a next step we then also measured $g_{vh}(t)$ for the same samples in 3D-DDLS mode (3D-DLS in VH geometry). For every detection angle $\theta$ we first determined the polarizer orientation manually using the same approach as already described for the experimental verification of eqn.~\ref{eq1}.  Note that no attempt was made here to also adjust the polarizer orientation for the incident beam for these experiments, but we will subsequently demonstrate below the negligible error that arises from this approach for these experiments. With the motorized polarizers, the variation of the polarizer tilt angle can of course be controlled remotely, rendering 3D-DDLS measurements of the angular dependence of $g_{vh}(t)$ much less time consuming.
Examples for the thus obtained correlation functions $g_{vh}(t)$ are shown in Fig.~\ref{fig:correlation} in red, and can be compared with the corresponding $g_{vv}(t)$ measured for the same systems and scattering angles. If the cross polarization of the instrument is set up properly, only a single exponential form of correlation function $g_{vh}(t)$ is expected for dilute samples. Only for a system close to a phase transition has a non-single exponential shape of the correlation function in VH been reported~\cite{kleshchanok2012dynamics}. The measured correlation functions obtained in VH geometry are indeed well-described by a single exponential given by Eqn.~\ref{eq2}, as demonstrated by the red lines in Fig.~\ref{fig:correlation}. From theses fits the relaxation rates $\Gamma$ in VH geometry are obtained and can be expressed as,
\begin{equation}
\Gamma = 6\:D{_R} + D{_T}\:q^{2}		
	\label{eq4}
\end{equation}
where D$_R$ is the rotational diffusion coefficient. In Fig.~\ref{fig:Dt_Dr}, the experimentally determined relaxation rates for VH are also plotted as $\Gamma$ vs q${^2}$ (red circles), and fitted with eq. \ref{eq4}. Both the VV and VH data show a clear q$^{2}$ dependence. Further, as expected from Eqns.~\ref{eq3} and \ref{eq4}, the VV and VH data are parallel, i.e. they have the same slope. \par

\begin{figure}[htb]
	\centering
		\includegraphics[width=0.48\textwidth]{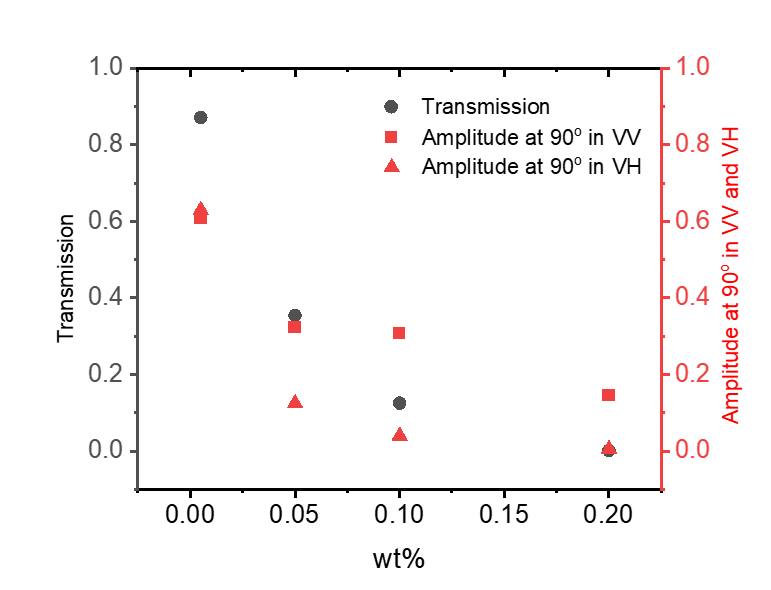}
	\caption{Transmission (circles) and amplitude of the correlation function for the VV (squares) and VH (triangles) geometry as a function of ellipsoid concentration.}
	\label{fig:transmission}
\end{figure}

\begin{figure}[htb]
	\centering
		\includegraphics[width=0.45\textwidth]{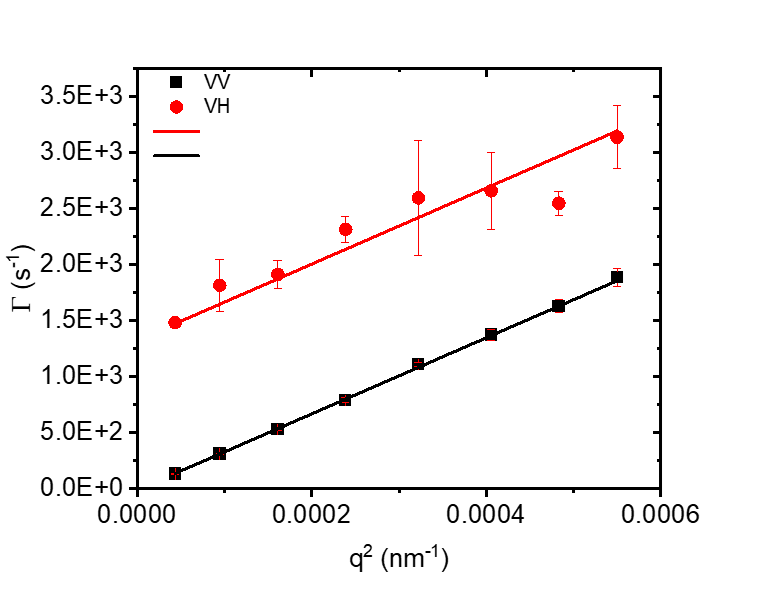}
	\caption{The extracted relaxation rate, $\Gamma$, as a function of q${^2}$ for VV (black) and VH (red) measurements for $\rho_{2}$ = 3.7 at a concentration of 0.35 wt\%, resulting in a transmission of 0.75 \%. Lines are fits to Eqn.~\ref{eq3} for VV (black) and Eqn.~\ref{eq4} for VH (red). }
	\label{fig:highc}
\end{figure}

While the strong multiple scattering and thus high turbidity and low transmission of our samples renders standard DLS and DDLS measurements impossible, with the Mod3D technology such measurements can still be performed. The 3D-DLS  cross-correlation technique isolates the single scattering contribution, while all contributions from multiple scattering enter the base line only and result in a decrease of the signal-to-base line ratio and thus the intercept of the correlation function. However, for samples with a transmission $T \gtrsim 10\%$ the quality of the correlation function data is still excellent and allows for accurate determination of the rotational and translational diffusion coefficients of the particles as demonstrated by the data shown in Fig.~\ref{fig:Dt_Dr}. \par

There is a direct relation between the transmission and the intercept of the correlation function, which is demonstrated in figure~\ref{fig:transmission}, where the measured transmission and the amplitude of the correlation functions at $\theta = 90^{\circ}$ are shown for the VV and VH mode, respectively. The decrease in intercept and thus in signal-to-noise increases the uncertainty in the values of $\Gamma$ in particular for the VH geometry. This is the result of a combination of a reduced intensity and thus a reduced signal to noise ratio for the depolarized scattering contributions as well as an enhanced proportion of multiple scattering due to the fact that multiple scattering also randomizes the polarization. At about a transmission of 1\%, the signal to noise ratio in the VH mode is so small that the resulting statistical error of the extracted relaxation rate $\Gamma$ becomes to large, while in VV mode somewhat lower transmission values can still be measured. This is further illustrated with data for particles with $\rho_2 = 3.7$ and a concentration of 0.35 wt\%. The sample has a transmission of 0.75 \%, and the corresponding $\Gamma$ vs $q^2$ plots for VV and VH are shown in Fig.~\ref{fig:highc}. Fig.~\ref{fig:highc} clearly shows that while the data quality for the VV measurements is still very high and allows for a quantitative determination of $D_T$ even at such a low transmission value, the individual values of $\Gamma$ obtained in VH carry a larger individual error, making a corresponding determination of $D_R$ (and $D_T$) also much more uncertain. \par  

\begin{figure}[hbt!]
	\centering
		\includegraphics[width=0.48\textwidth]{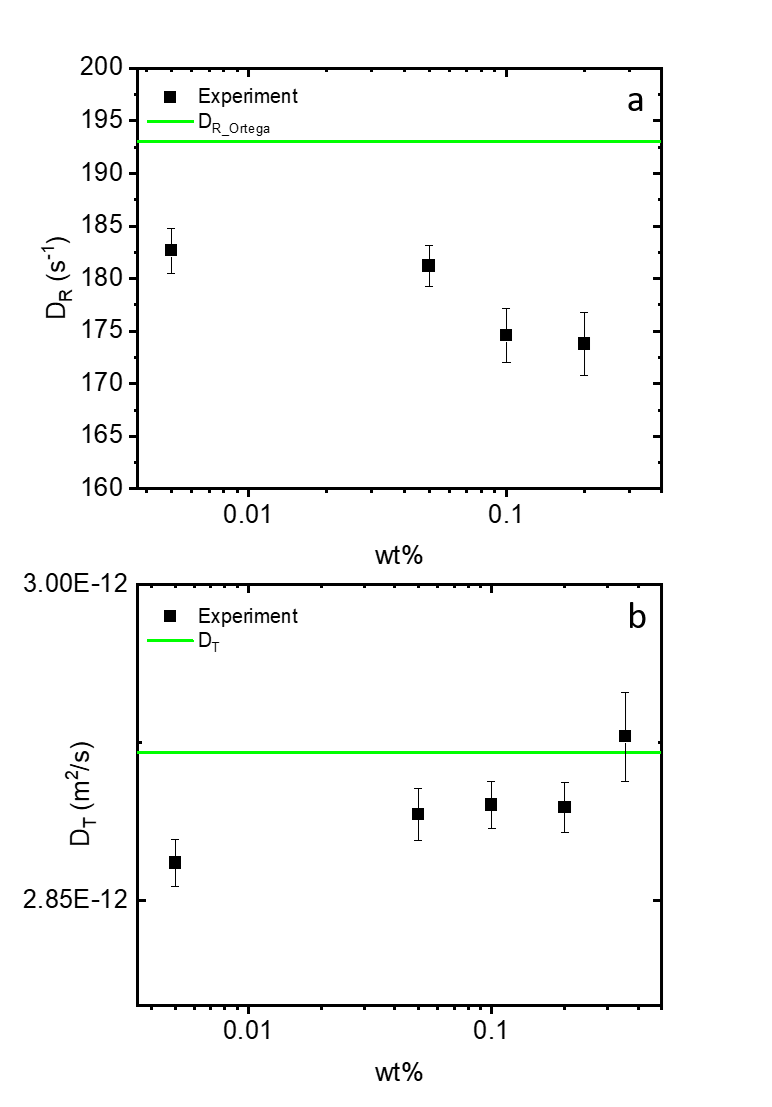}
	\caption{Extracted diffusion coefficients D$_{R}$ (a) and D$_{T}$ (b) (symbols) as a function of concentration for particles with aspect ratio $\rho_2$ = 3.7. The green line represents the theoretical prediction found in ref~\cite{ortega2003hydrodynamic}.}
	\label{fig:model}
\end{figure}

While almost all samples measured were highly turbid, the corresponding concentrations are very low, almost two orders of magnitude below the disorder to order transition or kinetic arrest~\cite{Pal2019}. Therefore interaction or crowding effects should not affect the diffusion coefficients measurably at these concentrations, and $D_T$ and $D_R$ should be independent of concentration, provided that multiple scattering contributions are properly suppressed. The experimentally obtained values of D$_T$ and D$_R$ are shown in Fig.~\ref{fig:model} for all concentrations for the aspect ratio $\rho_2$ (= 3.7). Both D$_T$ and D$_R$ are indeed independent of the concentration, which confirms that 3D-DLS using the mod3D technique is capable of reliably suppressing multiple scattering effects and allows us to correctly determine translational and rotational diffusion coefficients for highly turbid suspensions with transmission values as low as around 1\%. Using the TEM data as input, D$_T$ and D$_R$ were calculated using the theoretical approach of Ortega et al. for elongated particles, and the results are also shown as the green line in Fig~\ref{fig:model}. Given the known shape imperfections of these particles~\cite{Pal2022}, the agreement between the theoretical calculation and the measured values is surprisingly good. Note that two alternative expressions for different geometrical models, ellipsoids and spherocylinders, as described in Martchenko et al.~\cite{martchenko2011hydrodynamic}, resulted in significantly larger discrepancies between theoretical prediction and experimental results, with D$_T$ overestimated by 5-15\%, and even larger discrepancies for D$_R$.
 
As demonstrated by these measurements, 3D-DLS allows for precise measurements of translational and rotational diffusion coefficients for anisotropic particles even with samples that exhibit strong multiple scattering and thus high turbidity or low transmission. This is especially the case when combining this technique with the so-called modulated 3D cross-correlation technique (mod3D-DLS and mod3D-DDLS) that guarantees high amplitudes for the correlation function in the absence of multiple scattering. However, due to the 3D cross-correlation geometry measurements in VV and VH geometry require a relatively tedious measurement protocol, where the orientation of the polarizers and analyzers need to be adjusted for each scattering angle. We therefore also investigated the influence of contributions from multiple scattering as well as incorrect orientation of the polarizers on the outcome of the experiments. The results are summarized in Table 1. For this purpose a 0.1 wt\% samples of ellipsoids with an average aspect ratio of $\rho_2$ = 3.7 was used.\par

\begin{table*}[hbt]
\begin{ruledtabular}
\begin{tabular}{cccccccc}
		& VV-3D & VH-3D & VH-3D (90$^{\circ}$) & VH-3D (0$^{\circ}$) & VV-2D & VH-2D & VH-3D$^{\ast}$ \\ \hline
		&&&&&&& \\
 $D_{T}/D_{T,0}$&$1.0$&$1.0$ &$0.99$&$0.96$&$1.05$&$0.68$ &$1.0$\\
 $D_{R}/D_{R,0}$&$-$&$1.0$ &$0.99$&$1.12$&$-$&$1.16$ &$1.0$
\end{tabular}
\end{ruledtabular}
\caption{A comparison of the calculated normalized translational (D$_T/D_{T,0}$) and rotational (D$_R/D_{R,0}$) diffusion coefficients obtained for different geometries with a sample of particles with $\rho_2$ = 3.7 at a weight fraction of 0.1\% and a transmission of 19\%, where D$_{T,0}$ and D$_{R,0}$ are the correct values in the absence of effects from multiple scattering or interaction effects. VV-3D: measurement using mod3D-DLS with the correct adjustment of the polarizer orientation at each scattering angle. VH-3D: measurement using mod3D-DDLS with the correct adjustment of the polarizer orientation at each scattering angle. VH-3D(90$^{\circ}$): measurement using mod3D-DDLS with a polarizer tilt angle that was correctly determined at 90$^{\circ}$ scattering angle and then used for all the other scattering angles. VH-3D(0$^{\circ}$): measurement using mod3D-DDLS with a polarizer tilt angle that was determined with the direct beam (i.e., at 0$^{\circ}$ scattering angle) and then used for all scattering angles. VV-2D: standard DLS experiment in 2D with VV geometry. VH-2D: standard DDLS experiment in 2D with VH geometry. VH-3D$^{\ast}$: measurement using mod3D-DDLS with the correct adjustment of the polarizer orientation for the incident beam as well as at each scattering angle}
\end{table*}

We see clear and characteristic deviations for the different measurement protocols. The largest errors are obtained when performing DDLS experiments in standard 2D geometry, i.e. without suppressing contributions from multiple scattering (VV-2D and VH-2D). This leads to significant errors of up to 30\% in the translational and 16\% in the rotational diffusion coefficient. The use of the mod3D-DLS and mod3D-DDLS techniques lead to a significant reduction of the error, even without correctly adjusting for the correct tilt angle of the polarizers at each scattering angle independently. Due to our choice of the range of angles measured (30$^{\circ}\leq \theta \leq$ 135$^{\circ}$) and the symmetrical shape of the function describing the tilt angle $\omega$ around 90$^{\circ}$ scattering angle (Fig.~\ref{fig:polarization}), the effect is larger when initially adjusting the positions of the analyzers with the direct beam, i.e. at 0$^{\circ}$ scattering angle (VH-3D(0$^{\circ}$)), compared to a protocol where the position of the analyzer is determined with an isotropic scattering sample at a scattering angle of 90$^{\circ}$ (VH-3D(90$^{\circ}$)) and then used at all angles without further adjustment. However, it is important to point out that these results are specific for the chosen sample (optical anisotropy, range of scattering angles measured, degree of multiple scattering), and the only way to safely exclude artifacts in 3D-DDLS measurements originating from an incorrect alignment of the polarizers is to use the correction factor for the tilt angle given by eq. 1 for each angle measured.\par

The tilt angle of the polarizers is much less important for the VV geometry. Here, if the polarizers are not changed with angle $\theta$ according to eq. 1, a part of the polarized light will be blocked differently at each angle and consequently the intensity vs. scattering angle $\theta$ (or scattering vector $q$) will not be correct. This effect is however very small for the samples investigated, $<$0.1\%, over the whole angle range and can therefore be neglected. Secondly, by only turning the polarizers at the detectors, the incident vertical polarization and the horizontal polarization at the detectors are not in the same scattering plane. As a result, the scattering matrix is not nonzero for the non-eigenvectors. We experimentally turned both the incident polarizers and detector polarizers such that the polarization for both the incident and the detected scattered beam were in the same scattering plane (VH-3D*). We could see no difference between those measurements and the measurements where only the detector polarizers were turned (VH-3D).\par

\section*{Conclusions}
We have extended the established mod3D-DLS technique with its ability to suppress contributions from multiple scattering in DLS and SLS experiments to also perform depolarized dynamic light scattering measurements, and therefore to measure translational and rotational diffusion even for turbid samples. Due to the optical set-up of the 3D-DLS experiment, the direction of vertical and horizontal polarization of the scattered light changes with the scattering angle. While this has minor consequences for measurements using a VV geometry, it requires a readjustment of the orientation of the polarizers in front of the detectors at every investigated scattering angle in order to get the pure depolarized signal. While this renders the corresponding 3D-DDLS measurements tedious when performing them using manual adjustment of the polarizers, we have also built a motorized set-up that fully implements the required measurement protocols needed to perform such experiments. Using well-characterized colloidal model systems we have then tested this approach, and in particular investigated the performance as well as the magnitude of possible artefacts in case the measurement protocol would not be implemented completely. As a result, we now have a method that allows the conduction of DDLS experiments also with turbid samples, where multiple scattering would result in significant artefacts if standard DDLS experiments in 2D would be performed instead.

Using the 3D-DDLS set-up we have shown that it is possible to perform depolarized dynamic light scattering on samples with a transmission as low as 1\% with good accuracy and reproducibility. The current limitation comes from the increasing contribution from the multiple scattering, which decreases the intercept and thus the signal to noise ratio to a level, where in the end no reliable relaxation times can be extracted. The actual limit does of course depend on the system used, and turbid samples with a larger depolarization ratio would likely allow us to push the measuring limit to lower transmission values. This approach clearly opens up new possibilities to study dynamics of colloidal systems under non-ideal conditions. In the past, the characterization of the influence of interparticle interactions on rotational motion using DDLS was limited to systems that remained transparent up to very high packing fractions. With the extension of the mod3D crosscorrelation technique to DDLS experiments this requirement can now be relaxed, and we can also investigate systems that are not optically matched.

\begin{acknowledgments}
We gratefully acknowledge financial support from the European Union's Horizon 2020 research and innovation program through the European Soft Matter Infrastructure grant (731019-EUSMI). Crispin Hetherington is acknowledged for his help during TEM measurement.

\end{acknowledgments}
\appendix
\section*{Appendixes}
The detailed derivation of the change in the polarization angle for different scattering angles for the 3D-DLS set-up as presented in eq. 1 is given below. The 3D set-up has two beams that are focused on the scattering volume of the sample, see fig.~\ref{fig:schematic} main text. The two beams are mirror images of each other at the plane where the scattering volume is. Due to this the change in polarization angle for the two beams are the same but in opposite directions. Therefore, this derivation is the same for both the beams. The incident beams and the detectors lay in the same plane. One lens before the sample focuses the two beams at the scattering volume. A second lens after the sample makes the two beams parallel again before they are detected. The beams now come in an angle in the scattering volume  due to the scattering geometry and the scattering geometry of one of the beams can be described as a cone. The other beam creates the same scattering geometry mirrored in the center of the scattering volume. The direction of the “beam” along the cone can be described as a vector, $\vec{S_{c}}$:
\begin{equation}
\vec{S_{c}} = \begin{pmatrix} -\tan\alpha \cos\theta\\ -\tan\alpha \sin\theta\\ 1 \end{pmatrix}
	\label{eqA1}
\end{equation}
Where $\alpha$ is the cone angle (the angle of the laser is (90 - $\alpha$)) and $\theta$ is the rotational angle of the cone, i.e. scattering angle.\\
Normalizing the $\vec{S_{c}}$;
\begin{equation}
\left\|S_{c}\right\|= \frac{1}{\cos\alpha}
	\label{eqA2}
\end{equation}
Giving;
\begin{equation}
\hat{S_{c}}= \begin{pmatrix} -\sin\alpha \cos\theta\\ -\sin\alpha \sin\theta\\ \cos\alpha \end{pmatrix}
	\label{eqA3}
\end{equation}
Vector for the incident beam where $\theta$ = 0:
\begin{equation}
\hat{I_{0}}= \begin{pmatrix} -\sin\alpha\\0\\ \cos\alpha \end{pmatrix}
	\label{eqA4}
\end{equation}
and scattering ``beam":
\begin{equation}
\hat{I}_{S_{c}}= \begin{pmatrix} -\sin\alpha \cos\theta\\ -\sin\alpha \sin\theta\\ \cos\alpha \end{pmatrix}
	\label{eqA5}
\end{equation}
The vertical polarization of the scattering``beam", ($\hat{I}^{\bot}_{S_{c}}$) is normal to the ($\hat{I}_{S_{c}}$) at any $\theta$. To specify ($\hat{I}^{\bot}_{S_{c}}$), the plane at which ($\hat{I}^{\bot}_{S_{c}}$) is normal to has to be determined. This plane is determined by ($\hat{I}_{S_{c}}$) and ($\vec{I}_{tan}$), where ($\vec{I}_{tan}$) is the tangent vector of the cone circle. 
The line that is the tangent of a circle:
\begin{equation}
x\cdot x_0+y\cdot y_0=r^2
	\label{eqA6}
\end{equation}
where $x_0$ and $y_0$ is the point of the circle and r is the radius of the cone at a distance z from the top of the cone:
\begin{equation}
	r^2 = (-z\:tan\alpha)^2
	\label{eqA7}
\end{equation}
The vector along the tangent, ($\vec{I}_{tan}$), is:
\begin{align*}
	\vec{I}_{tan}& = \begin{pmatrix} x_0\\ y_0\\ z \end{pmatrix} - \begin{pmatrix} x\\ y\\ z \end{pmatrix}\\
		x_0 & = - z \tan\alpha \cos\theta  \\
	  y_0 & = - z \tan\alpha \sin\theta   \\
		y & = \frac{(-z \tan\alpha)^2 + z \tan\alpha \cos\theta \: x}{- z \tan\alpha \sin\theta} \\
		\phantom{y} & = - \frac{z \tan\alpha + \cos\theta \: x}{\sin\theta}
	\label{eqA8}
\end{align*}
Thus,
\begin{align}
	\vec{I}_{tan} & = \begin{pmatrix} - z \tan\alpha \cos\theta\\ - z \tan\alpha \sin\theta\\ z \end{pmatrix} - \begin{pmatrix} x\\ - \frac{z \tan\alpha + \cos\theta \: x}{\sin\theta}\\ z \end{pmatrix}\\
	& = \begin{pmatrix} - (z \tan\alpha \cos\theta + x) \\ \frac{\cos\theta}{\sin\theta}(z \tan\alpha \cos\theta + x)\\ 0 \end{pmatrix}
	\label{eqA9}
\end{align}
Normalizing ($\vec{I}_{tan}$) gives:
\begin{equation}
\hat{I}_{tan}= \begin{pmatrix} -\sin\theta\\\cos\theta\\ 0 \end{pmatrix}
	\label{eqA10}
\end{equation}
The vector normal to the plane, $\hat{I}^{\bot}$, defined by $\hat{I}_{S_{c}}$ and $\vec{I}_{tan}$ is:
\begin{equation}
\hat{I}^{\bot} = \hat{I}_{S_{c}} \times \hat{I}_{tan} = \begin{pmatrix} -\sin\alpha \cos\theta\\ -\sin\alpha \sin\theta\\ \cos\alpha \end{pmatrix} \times \begin{pmatrix} -\sin\theta\\\cos\theta\\ 0 \end{pmatrix}
	\label{eqA11}
\end{equation}
The vector that would correspond to the vertical component of the of the cone at the incident beam ($\theta$=0), ($\hat{I}^{\bot}_{0}$) and the ``beam" at the detector ($\theta=\theta$), ($\hat{I}^{\bot}_{S_{c}}$) would have the vector representation in the cone geometry as:
\begin{eqnarray}
\hat{I}^{\bot}_{S_{c}} & = \begin{pmatrix} \cos\alpha \cos\theta\\ \cos\alpha \sin\theta\\ \sin\alpha \end{pmatrix} \nonumber \\
\hat{I}^{\bot}_{0} & = \begin{pmatrix} \cos\alpha\\0\\\sin\alpha \end{pmatrix} \nonumber
\end{eqnarray}
To find out how much the scattering polarization deviates from vertical polarization at the detection due to the cone geometry when the incident beam is vertically polarized, we project $\hat{I}^{\bot}_{0}$ on $\hat{I}_{S_{c}}$:
\begin{equation}
\vec{I}^{s}_{proj} = \frac{\left(\hat{I}^{\bot}_{0} \cdot \hat{I}_{S_{c}}\right)}{\left(\left\|{I}_{S_{c}}\right\|\right)^2} \hat{I}_{S_{c}} 
	\label{eqA12}
\end{equation}
Since $\left\|{I}_{S_{c}}\right\|^2$ = 1,
\begin{eqnarray}
\vec{I}^{s}_{proj} = &&\begin{pmatrix} \cos\alpha\\0\\\sin\alpha \end{pmatrix} \cdot \begin{pmatrix} -\sin\alpha \cos\theta\\ -\sin\alpha \sin\theta\\ \cos\alpha \end{pmatrix} \hat{I}_{S_{c}} \nonumber\\
\phantom{\vec{I}^{s}_{proj}} = &&\sin\alpha \cos\alpha (1 - \cos\theta) \hat{I}_{S_{c}}
	\label{eqA13}
\end{eqnarray}
The normal to $\hat{I}_{S_{c}}$, $\vec{I}^{\bot}_{proj}$ of this projection of the vertical component of the incident beam:
\begin{eqnarray}
\vec{I}^{\bot}_{proj} = && \hat{I}^{\bot}_{0} - \hat{I}^{s}_{proj} \nonumber\\
\phantom{\vec{I}^{\bot}_{proj}} = &&\begin{pmatrix} \cos\alpha\\0\\\sin\alpha \end{pmatrix} - \sin\alpha \cos\alpha (1 - \cos\theta) \begin{pmatrix} -\sin\alpha \cos\theta\\ -\sin\alpha \sin\theta\\\cos\alpha \end{pmatrix} \nonumber
\end{eqnarray}
\begin{equation}
\vec{I}^{\bot}_{proj} =  \begin{pmatrix} \cos\alpha\:(1 + \sin^2 \alpha \cos\theta\:(1 - \cos\theta))\\\sin^2 \alpha \cos\alpha \sin\theta\:(1 - \cos\theta)\\\sin\alpha - \sin\alpha \cos^2 \alpha\:(1 - \cos\theta) \end{pmatrix} 
	\label{eqA14}
\end{equation}
The angle between $\vec{I}^{\bot}_{proj}$ and $\hat{I}^{\bot}_{S_{c}}$ gives the turning of the polarization due to the cone geometry:
\begin{equation}
\cos\omega = \frac{\vec{I}^{\bot}_{proj} \cdot \hat{I}^{\bot}_{S_{c}}}{\left\|{I}^{\bot}_{proj}\right\| \left\|{I}^{\bot}_{S_{c}}\right\|}  
	\label{eqA15}
\end{equation}
where $\omega$ is the turning of the polarization. With $\left\|{I}^{\bot}_{S_{c}}\right\|$ = 1, 
\begin{equation}
\left\|{I}^{\bot}_{proj}\right\| = \sqrt{1 - \sin^2\alpha \cos^2 \alpha\:(1 - \cos\theta)^2}
	\label{eqA16}
\end{equation}
\\
\begin{widetext}
\begin{eqnarray}\label{eqA17}
\vec{I}^{\bot}_{proj} \cdot \hat{I}^{\bot}_{S_{c}} & = \begin{pmatrix} \cos\alpha\:(1 + \sin^2 \alpha \cos\theta\:(1 - \cos\theta))\\\sin^2 \alpha \cos\alpha \sin\theta\:(1 - \cos\theta)\\\sin\alpha - \sin\alpha \cos^2 \alpha\:(1 - \cos\theta) \end{pmatrix} \cdot \begin{pmatrix} \cos\alpha \cos\theta\\ \cos\alpha \sin\theta\\\sin\alpha \end{pmatrix} 
\phantom{\vec{I}^{\bot}_{proj} \cdot \hat{I}^{\bot}_{S_{c}}} 
= \cos^2 \alpha \cos\theta + \sin^2 \alpha 
\end{eqnarray}
\end{widetext}
This gives the turning of the polarization, $\omega$, as a function of scattering angle, $\theta$, and the incident beam due to the focusing, $\alpha$:
\begin{equation}
\omega = \cos^{-1}\left(\frac{\cos^{2}\alpha \;\cos\theta\;+\;\sin^{2}\alpha}{\sqrt{1-\sin^{2}\alpha\:\cos^{2}\alpha \:(1 - \cos\theta)^{2}}}\right)		
\end{equation}

\bibliography{ref_3DDLS}

\end{document}